# Temporal Spectral Noise-Floor Adaptation for Error-Intolerant Trigger Integrity in IoT Mesh Networks

Sergii Makovetskyi and Lars Thomsen

***Abstract*** **In this paper, we present a lightweight, embedded algorithm for autonomous edge event triggering in IoT sensor nodes suitable for operating in mesh networks. The device acquires local sensor data, performs deterministic FFT spectral feature extraction in firmware, and maintains a temporal spectral noise-floor baseline that absorbs non-stationary environmental excitations such as rain, wind, and mechanical vibration. While adaptive thresholds in IoT sensor nodes are often applied to manage communication load or stabilize long-term metrics, this work focuses on maintaining a time-evolving spectral noise floor to preserve event trigger reliability in dynamic environments. Our method targets trigger integrity under environmental non-stationary conditions, enabling calibration-free deployment of autonomous nodes; without shared noise models or cloud-side inference. Local decision authority preserves node responsiveness when connectivity is intermittent and mitigates security risks inherent in centralized remote-analysis systems. We validate the algorithm in a single node mesh sensor deployed in a dynamic outdoor environment using a radar-class proximity sensor as one example sensor modality. Results demonstrate substantial suppression of nuisance-induced triggers, reduced false-event traffic amplification in the mesh, bounded embedded execution, and reliable detection sensitivity to true spectral signatures.**

## I. INTRODUCTION

THE rapid proliferation of distributed IoT sensor nodes in unstructured outdoor environments has exposed persistent system-design challenges at the intersection of communication efficiency, detection reliability, and autonomous edge processing outlined by Pioli et al. [1] and Sadri et al. [2]. Data transfer scales proportional with the number of nodes in a sensor network with concomitant performance limitations caused by network congestion, transmission latency and bandwidth constraints. The combined effects compromise the responsiveness required for time-critical applications [3],[4]. Two dominant research directions have emerged to address these challenges:

The first focuses on *communication efficiency through adaptive data* reduction, where thresholds are dynamically adjusted at sensor nodes to minimize uplink transmissions while preserving reconstruction fidelity; whereas the second pursues *spectral anomaly detection at the edge*, where embedded FFT processing enables transform-domain feature extraction on resource-constrained hardware.

An example of the first approach is described by Hussein et al. [5] who demonstrated that distributed prediction and compression schemes can reduce data transmission by up to 96% while maintaining acceptable accuracy. Similarly, Idrees and Al-Qurabat [6] proposed a protocol with an initial data reduction of redundant data at the sensor level, and a second data reduction by removal of similar data sets received from various sensors at the fog nodes before transmission to the base station.

These approaches operate in the time domain typically detecting drift or applying deep learning to generate adaptive sampling rates to reduce network load [7],[8]. These approaches prioritize reduction in data load in the network over trigger integrity and can lead to an unpredictable amount of false negative rates in detection-critical applications where the risk is that the deep learning model fails and an event occurs outside the models predicted time domains.

An example of the second approach is given by Bhoi et al. [9] who demonstrated Short-Time Fourier Transform (STFT) feature extraction on edge devices for grid power quality monitoring, triggering cloud transmission only when spectral anomalies are detected; and by Xie et al. [10] who proposed binary-convolution networks that encode time-series data into spectral features at the edge, achieving 96% data reduction while maintaining anomaly detection accuracy.

The feasibility of FFT-based processing on microcontrollers has been further validated in industrial condition monitoring applications [11],[12]. However, these spectral approaches typically employ fixed detection thresholds or cloud-side classifiers which renders them susceptible to environmental non-stationarity fluctuations in background noise caused by wind, rain, mechanical vibration, or platform motion. The combined effect of all variable noise sources can lead to an elevation in the false positive rates when thresholds remain static.

Despite significant advances in both domains, a critical gap persists: *no existing work combines FFT-based spectral feature extraction with temporal adaptive noise-floor tracking for autonomous trigger decisions at the mesh node level*. Current adaptive threshold methods operate predominantly in the time domain and target data reduction rather than detection integrity [13], [14]. Whereas spectral anomaly detectors assume either

Date of submission: 2nd of February, 2026, This work was supported in part by Gnacode Inc., a Canadian biotech company, with its registered address in T1B4S3 Medicine Hat, Alberta, Canada under a grant 1037487 by the Canadian Research Counsil. *Corresponding author: Lars Thomsen.*

Lars Thomsen, M.Sc., Ph.D. Author is the managing director of Gnacode Inc (e-mail: lt@gnacode.com). Sergii Makovetskyi, M.Sc., Eng. Author is a PhD student at Kharkiv National University of Radio Electronics, Nauky Ave. 14, Kharkiv, 61166, Ukraine (email: serhii.makovetskyi@nure.ua)

centralized interpretation or static environmental baselines that fail under non-stationary conditions. Furthermore, existing edge-cloud architectures typically require bidirectional communication for threshold adaptation or model updates as described by GabAllah et al., [15] and by Zhu and Xie, [16], introducing latency that is unacceptable for safety-critical trigger applications where immediate response is essential. This paper addresses this gap by presenting a system design that integrates spectral feature extraction with autonomous adaptive thresholding entirely within embedded MCU firmware. Our approach maintains a *temporal spectral noise-floor baseline* at each node represented as a time-evolving estimate of environmental spectral energy that absorbs non-stationary excitations such as rain, wind, and mechanical vibration. Unlike spatial CFAR (Constant False Alarm Rate) methods used in radar signal processing, which estimate noise from neighbouring range cells, our method tracks noise evolution temporally within each frequency bin, enabling the threshold to adapt continuously to changing environmental conditions without requiring calibration or cloud-side feedback.

The key insight is that environmental disturbances and true signal events occupy distinct positions in the spectral-temporal space: environmental noise elevates spectral energy gradually and diffusely across frequency bins, while genuine events produce sharp, localized spectral deviations that exceed the adapted noise floor. By implementing this discrimination entirely at the edge, each sensor node can make autonomous, verified trigger decisions with high integrity. An additional benefit is that mesh network traffic reduces to minimal event logs consisting of, e.g., delta-encoded timestamps for actual events. The high-integrity detection and minimal event logs also reduce false event traffic that in noisy situations can cause network congestion and destabilize multi-hop IoT topologies.
The main contributions of this work are:

1. A formalized temporal spectral noise-floor tracking algorithm implemented entirely in embedded MCU firmware, enabling calibration-free deployment across heterogeneous environmental conditions.

2. A deterministic trigger integrity rule operating on spectral FFT features that simultaneously minimizes false positives (through adaptive noise-floor tracking) and false negatives (through spectral event detection), without relying on cloud classifiers or spatial averaging.

3. Demonstration of nuisance-resilient event triggering in dynamic outdoor environments, showing substantial suppression of environmentally induced false triggers while maintaining high detection sensitivity to true spectral signatures.

4. Demonstrated mesh network traffic reduction of 98% achieved through local decision authority, preventing cascading false-event propagation in multi-hop IoT deployments.

This positions our work at the intersection of time-domain adaptive thresholding for network efficiency and FFT-based edge anomaly detection; with a clear objective of achieving robust event trigger integrity under dynamically changing environmental spectral interference in autonomous IoT mesh nodes.

The remainder of this paper is organized as follows. Section II reviews related work in adaptive data reduction and edge-based anomaly detection. Section III presents the system architecture and theoretical framework for temporal spectral noise-floor adaptation. Section IV describes experimental validation using a single node sensor deployment. Section V is discussion that concludes with a summary and directions for future work.

## II. RELATED WORK

This section reviews the two principal research streams relevant to autonomous event triggering in IoT sensor networks: adaptive thresholding for communication efficiency and spectral feature extraction for edge anomaly detection. We then examine the challenges posed by environmental non-stationarity and identify the research gap addressed by this work.

### *A. Adaptive Thresholding for Communication Efficiency*

The fundamental tension between data fidelity and network efficiency in IoT sensor deployments has motivated extensive research into adaptive data reduction techniques. Pioli et al. [1] conducted a systematic mapping of edge-based data reduction, categorizing approaches into compression, prediction, and aggregation methods. A common theme across these approaches is the use of adaptive thresholds to gate transmissions based on signal change detection.
Zhang et al. [17] proposed an autonomous data reduction method that dynamically adjusts thresholds based on variance between consecutive sensor readings. By transmitting only when values deviate substantially from predictions, their approach achieved up to 80% data reduction while maintaining reconstruction accuracy. However, this time-domain scheme relies on decimation, selective omission of samples, which inherently risks missing transient events that occur between transmission intervals.
Prediction-based dual schemes represent another class of adaptive methods. Hussein et al. [18] combined autoregressive prediction with adaptive piecewise constant approximation to achieve 85–96% transmission reduction. Wu et al. [8] demonstrated cloud-edge collaborative prediction where edge nodes transmit only when predicted values deviate beyond tolerance thresholds, reducing transmissions by 88% while maintaining deviation bounds. Similarly, Putra et al. [7] deployed deep learning prediction on base stations, achieving 33% energy reduction through selective transmission.
Fog-based aggregation architectures extend adaptive thresholding across network tiers. Idrees and Al-Qurabat [6] proposed dynamic time warping clustering at fog nodes to eliminate redundant data sets before gateway transmission.

Sadri et al. [2], [19] surveyed fog data management approaches, identifying threshold-based filtering as the predominant technique for reducing data volume in hierarchical IoT architectures.

Despite their effectiveness for bandwidth reduction, these time-domain adaptive methods share a fundamental limitation in that they optimize for communication efficiency rather than detection integrity. When the primary objective is triggering sensor events rather than reconstructing them the missed detections due to decimation can be catastrophic in safety-critical applications.

### *B. Spectral Feature Extraction for Edge Anomaly Detection*

Transform-domain processing offers an alternative paradigm that extracts discriminative features from sensor signals before making detection decisions. The feasibility of embedded FFT processing on resource-constrained microcontrollers has been demonstrated across multiple application domains.

Arciniegas et al. [20] presented a TinyML system for motor bearing fault detection that integrates FFT-based spectral analysis with deep learning classification on an ESP32 microcontroller. Their system achieved 96.5% accuracy with 300ms latency from data acquisition to alert generation, validating that sophisticated spectral processing is achievable on edge devices. Trilles et al. [21] conducted a systematic mapping of AIoT anomaly detection, finding that classification algorithms, in particularly convolutional neural networks operating on spectral features, dominate recent embedded implementations.

Hammad et al. [22] demonstrated unsupervised TinyML for urban noise anomaly detection, employing spectral features to identify acoustic events without labeled training data. Their approach highlights the advantage of frequency-domain representation for distinguishing structured events from background noise. Bhoi et al. [9] applied Short-Time Fourier Transform (STFT) feature extraction at the edge for grid power quality monitoring, triggering cloud transmission only when spectral anomalies are detected and achieved a reduction in cloud storage requirements by 95%.

Binary feature encoding represents an emerging approach to bridge spectral analysis and network efficiency. Xie et al. [23] proposed encoding time-series data into binary spectral vectors at edge nodes, achieving 96% data reduction while maintaining anomaly detection performance. Their hierarchical temporal memory detector operates on compressed features transmitted to the cloud, demonstrating that spectral signatures can be preserved through aggressive dimensionality reduction.

Edge-cloud architectures for anomaly detection have been explored extensively. Hafeez et al. [3] surveyed data handling strategies for industrial IoT (IIoT) predictive maintenance, comparing device-edge, edge-cloud, and federated approaches. They identify a critical trade-off: edge-side processing reduces latency but requires lightweight models, while cloud-side inference offers higher accuracy at the cost of transmission delay. GabAllah et al. [15] proposed intelligent data filtering at the edge to reduce inter-tier communication by 50% while maintaining comparable classification performance.

However, existing spectral anomaly detection approaches predominantly employ either fixed thresholds or cloud-trained classifiers. Fixed thresholds fail under environmental non-stationarity, while cloud-dependent classification introduces latency incompatible with real-time triggering requirements.

### *C. Environmental Non-Stationarity in Outdoor Deployments*

Outdoor IoT sensor deployments face unique challenges from non-stationary environmental excitations that corrupt baseline signal characteristics. Unlike controlled industrial environments where background noise is relatively stable, outdoor sensors must contend with wind, rain, mechanical vibration, thermal drift, and if mounted in mobile platforms they face the vehicle or drone frame excitation as an additional noise source.

The radar signal processing community has long addressed analogous problems through Constant False Alarm Rate (CFAR) detection, which estimates noise power from spatial reference cells surrounding the cell under test. However, spatial CFAR faces two failure modes in IoT deployments. First, when environmental disturbances propagate across a sensor network (e.g., a passing weather front), leading-edge sensors experience elevated noise while their reference cells remain unaffected, causing false alarms until the disturbance reaches neighboring nodes. Second, when localized interference affects a single sensor (e.g., mechanical vibration on a drone-mounted node), the clean neighboring cells produce an artificially low noise estimate, again triggering false alarms. Both scenarios violate CFAR's assumption of spatial homogeneity within the reference window [24], [25].

Adaptive noise estimation in IoT contexts has received limited attention. Oikonomou et al. [4] surveyed intelligent models for edge mesh challenges, identifying environmental adaptability as a key unsolved problem for autonomous sensor networks. Sittón-Candanedo and Corchado [26] emphasized that edge computing's primary value lies in pre-processing data under local conditions before cloud transmission, yet practical methods for adapting to changing environmental baselines remain underdeveloped.

The consequence of static thresholds under environmental non-stationarity is two-fold: elevated noise floors cause excessive false positives, while conservative threshold settings to avoid false alarms result in missed detections (false negatives). Neither outcome is acceptable in applications requiring high trigger integrity.

### *D. Research Gap*

The reviewed literature reveals a clear gap at the intersection of three requirements: (1) spectral feature extraction for robust event discrimination, (2) adaptive noise-floor tracking for environmental resilience, and (3) autonomous edge decision-making for latency-critical triggering.

Existing adaptive threshold methods operate primarily in the time domain and optimize for network efficiency rather than detection integrity [17],[18]. Spectral anomaly detectors validate FFT-based feature extraction on embedded hardware but rely on fixed thresholds or cloud-side inference [20],[23]. No prior work maintains a temporal spectral noise-floor baseline that evolves with environmental conditions while enabling autonomous trigger decisions entirely at the mesh node level.

This paper addresses this gap by proposing a system that tracks spectral magnitude over time within each frequency bin, automatically adapting detection thresholds to absorb environmental excitations while preserving sensitivity to genuine event signatures. The approach operates entirely in firmware, requires no calibration or cloud feedback, and produces minimal trigger outputs suitable for low-overhead mesh propagation.

## III. SYSTEM ARCHITECTURE AND THEORETICAL FRAMEWORK

This section presents the theoretical foundations of temporal spectral noise-floor adaptation for autonomous edge triggering. We first describe the overall system architecture, then formalize the spectral feature extraction, noise-floor tracking algorithm, and integrity-gated trigger decision rule.

### *A. System Architecture Overview*

The proposed system implements a three-stage processing pipeline entirely within embedded MCU firmware at each sensor node, as illustrated in Fig. 1(C). The pipeline consists of:

1. *Spectral Feature Extraction*: Raw sensor samples are transformed to the frequency domain via FFT, producing magnitude estimates for frequency bins of interest.
2. *Temporal Noise-Floor Tracking:* An adaptive baseline tracks the spectral magnitude in each frequency bin over time, absorbing gradual environmental variations while remaining responsive to sudden deviations.
3. *Integrity-Gated Trigger Decision:* A deterministic rule compares current spectral magnitudes against the adapted noise floor, issuing trigger events only when deviations exceed configured margins.

The output is a binary trigger indicator with minimal metadata (timestamp, trigger strength), suitable for low-overhead propagation through opportunistic-routed mesh networks. This architecture differs fundamentally from transmission-gated adaptive threshold systems as described by Zhang et al. [17], which reduce data volume but still require downstream processing to detect events, and from cloud-dependent spectral classifiers described by Xie et al. [23], which introduce latency incompatible with real-time triggering. Fig. 1 illustrates the comparative behavior of three detection paradigms. Approach (A) employs temporal adaptive thresholding with decimation to reduce network load; while effective at suppressing false positives, decimation causes missed events when transients occur between transmitted samples. Approach (B) applies FFT-based anomaly detection with fixed thresholds, achieving rapid edge decisions but suffering excessive false positives when environmental noise elevates the spectral floor. The proposed method (C) combines FFT-based spectral feature extraction with adaptive noise-floor tracking, achieving both low false-negative rates (through continuous spectral monitoring) and low false-positive rates (through temporal baseline adaptation).

### *B. Spectral Feature Extraction via Embedded FFT*

Let x[n] denote a sequence of N samples acquired from the sensor at sampling rate $f_s$. The Discrete Fourier Transform (DFT) maps this time-domain sequence to the frequency domain described:

$$X[k] = \sum_{n=0}^{N-1} x[n]\, e^{-j2\pi kn/N}, \quad k = 0,1,\ldots,N-1 \tag{1}$$

where $X[k]$ represents the complex-valued spectral coefficient at frequency bin $k$, corresponding to frequency $f_k = k \cdot f_s/N$. The spectral magnitude is computed as:

$$|X[k]| = \sqrt{Re(X[k])^2 + Im(X[k])^2} \tag{2}$$

For computational efficiency on resource-constrained MCUs, we employ the Cooley-Tukey radix-2 Fast Fourier Transform algorithm, which reduces complexity from $O(N^2)$ to $ON\log(N)$ [27]. With a frame sizes of $N = 128$ samples, FFT computation completes within milliseconds on modern ARM Cortex-M class processors [25],[28]. Rather than retaining all $N/2$ unique frequency bins, we define a feature vector comprising bins of interest:

$$f_t = [|X[k_1]|, |X[k_2]|, \ldots, |X[k_M]|]^T \tag{3}$$

where

$$k_1, k_2, \ldots, k_M$$

are selected based on the expected spectral signature of target events. This reduction from $N/2$ bins to $M$ features gives a further reduction in computational and memory requirements while focusing detection sensitivity on relevant frequency bands.

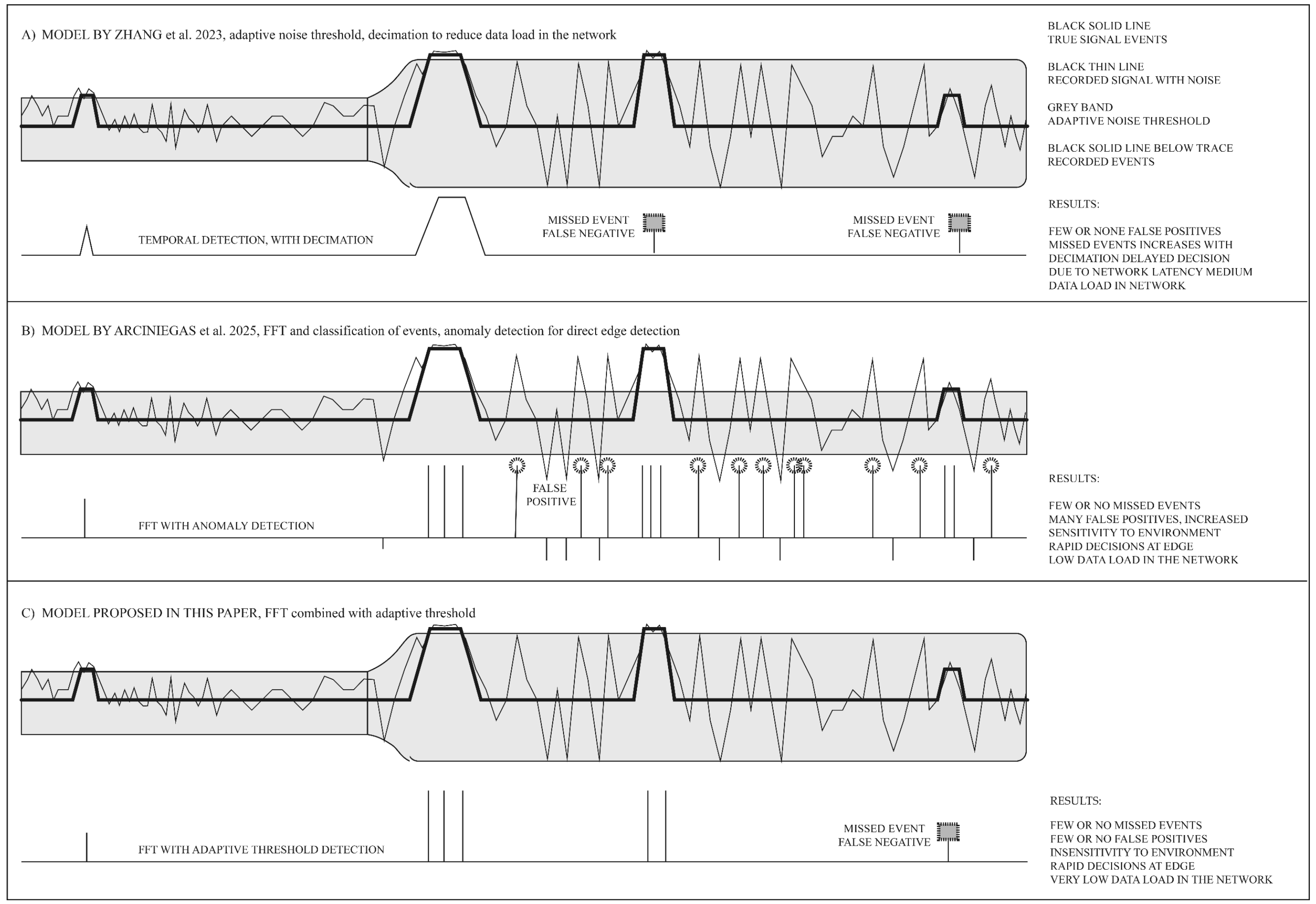

**Fig.1.** Comparison of three event detection paradigms: (A) temporal adaptive thresholding with decimation Zhang et al. [17] , (B) FFT-based anomaly detection with fixed threshold by Arciniegas et al. [20] , and (C) the proposed FFT-based detection with adaptive spectral noise-floor tracking. Solid black lines indicate true signal events; grey lines show recorded signal with noise; grey bands represent detection thresholds. The resulting detected events are shown in a solid line below with stipulated circles marking false positives, and stipulated squares marking false negatives. The proposed model has no false positives and no or a very low rate of false negatives.

*Spectral representation advantage:* The key insight motivating spectral processing is that genuine events and environmental noise occupy distinct regions in the frequency-magnitude space. True events typically produce sharp, localized spectral peaks at characteristic frequencies, while environmental excitations (wind, rain, vibration) tend to elevate energy diffusely across broader frequency ranges. This separation enables more robust discrimination than time-domain amplitude thresholding alone.

### *C. Temporal Spectral Noise-Floor Baseline Tracking*

The core contribution of this work is a noise-floor tracking algorithm that maintains a time-evolving estimate of background spectral magnitude in each frequency bin. Unlike spatial CFAR methods used in radar [24], [25], which estimate noise from neighboring spatial cells under assumptions of homogeneity, our approach tracks noise evolution temporally within each frequency bin independently.

#### *Dual-Stage Cascaded Median Filter Architecture*

For each frequency bin $k$ of interest, we maintain a noise-floor estimate using a two-stage cascaded median filter architecture. The median filter, introduced by Tukey [29] for robust nonlinear smoothing, returns the middle value of a sorted window and provides inherent resistance to outliers which is a property formalized by Gallagher & Wise [30] who proved that median filters preserve locally monotone structure while rejecting impulsive contamination.

The noise-floor estimate $\hat{N}_k[t]$ is computed as:

$$\hat{N}_k[t] = \text{median}_\gamma\{|X_k[t-i]|\}_{i=0}^{\gamma-1} \tag{4}$$

where the two stages serve distinct purposes:

*Stage 1, Digital Noise Suppression* ($\gamma = \gamma_d \in [2, 5]$): A fast median filter with small window size removes high-frequency digital artifacts including ADC quantization noise, bit-scaling transients, and computational rounding errors. This exhibits frame-to-frame variation with correlation times shorter than the FFT frame period effectively suppresses fast transient artifacts and introduces a minimal latency of 2 to 5 frames.

*Stage 2, Analog Drift Tracking* ($\gamma = \gamma_a \in [64, 128]$): A slow median filter with large window size tracks gradual environmental variations such as thermal drift, component aging, and slow-changing ambient conditions, and therefore provides a stable baseline that tracks these slow variations without responding to transient events.

*Median Filter Implementation*

Each median filter stage maintains a circular buffer and computes the median via in-place sorting:

$$\text{median}_{\gamma}\{x[t], x[t-1], \ldots, x[t-\gamma+1]\} = x_j, j = \left\lceil \frac{\gamma}{2} \right\rceil \quad (5)$$

where $x_{(j)}$ denotes the *j*-th order statistic (the *j*-th smallest value) of the window contents. For embedded implementation, direct sorting suffices for small windows ($\gamma_d \leq 5$), while partial sorting improves efficiency for larger windows ($\gamma_a$).

*Robustness Properties*

The median filter possesses a 50% breakdown point [29], meaning up to half the samples in the window can be arbitrarily corrupted without affecting the output. This robustness is critical for noise-floor estimation because:

1. *Genuine events do not corrupt the baseline*: Even sustained events occupying up to $\lfloor\gamma/2\rfloor - 1$ consecutive frames leave the median unchanged.
2. *Impulsive interference is rejected:* Sporadic high-energy artifacts are automatically excluded from the baseline estimate.
3. *No explicit censoring required:* The median filter's inherent order-statistic selection provides implicit outlier rejection.

*D. Threshold-Based Trigger Decision Rule*

Given the current spectral magnitudes $|X_k[t]|$ and adapted noise floors $\hat{N}_k[t]$ from the dual-stage median filter, the trigger decision rule determines whether an event has occurred. We formulate this as a multiplicative threshold test:

$$D_k[t] = \{1 \; if \; |X_k[t]| > \zeta_k \cdot \hat{N}_k[t]; 0 \text{ otherwise}\} \quad (6)$$

where $\zeta_k \in [1, 2]$ is a per-bin threshold coefficient. The multiplicative formulation scales the detection margin proportionally to the ambient noise level, maintaining consistent sensitivity across varying environmental conditions. A coefficient of $\zeta_k = 1.5$, requires the spectral magnitude to exceed the adapted noise floor by a factor of 1.5 before triggering detection.

*Single-Bin Trigger Architecture*

The system employs a single-bin trigger architecture: a system-level event E[t] is declared when any monitored frequency bin exceeds its threshold:

$$E[t] = \{1 \text{ if any } D_k[t] = 1, k \in K; 0 \text{ otherwise }\} \quad (7)$$

where $K$ is the set of monitored frequency bins. This approach maximizes detection sensitivity and genuine events producing elevated magnitude in characteristic frequency bands are immediately captured without requiring corroboration across multiple bins.

*Double-Buffered Processing Architecture*

To enable continuous monitoring without gaps, the system employs a double-buffered architecture with separate sampling and analysis buffers:

*Sampling buffer:* Continuously acquires sensor samples at rate $f_s$, filling the buffer with $N$ samples per frame.

*Analysis buffer:* Contains the previously acquired frame, undergoes FFT computation, noise-floor update, and trigger evaluation while the sampling buffer fills.

*E. Impact on Mesh Network Data Load*

The proposed architecture fundamentally changes the nature of data propagated through the mesh network. Rather than transmitting raw samples, reduced samples, or extracted features for remote processing, each node transmits only *binary trigger indicators* $E[t]$ with minimal metadata.

*Payload Comparison*

Table I compares payload characteristics across different architectural approaches:

**TABLE I: Payload Comparison**

| Approach | Payload Content | Size per Event | Decision Latency |
|---|---|---|---|
| Raw streaming | Time-domain samples | $N \times B$ = 2048 bits | Cloud-dependent |
| Decimated streaming | Reduced samples | $(N/D) \times B = 512$ bits | Cloud-dependent |
| Feature transmission | Spectral magnitudes | $M \times B_f$ bits = 256 bits | Cloud-dependent |
| Proposed | Trigger flag + metadata | ~32-64 bits | Local (<3 ms) |

where $N$ is frame size, $B$ is bits per sample, $D$ is decimation factor, M is number of monitored frequency bins, and $B_f$ is bits per feature. For typical values ($N$=128, $B$=16, $D$=4, $M$=16, $B_f$=16), the proposed approach transmits 32-64 bits per event compared to 2048 bits for raw streaming, 512 bits for decimated streaming, and 256 bits for feature transmission which is a reduction in network load of 4-64 times depending on the baseline approach.

*False-Event Traffic Amplification*

In multi-hop mesh networks, false triggers have compounding effects: each false event consumes bandwidth at every hop between source and sink, potentially triggering retransmissions and congestion backoff. We define the *traffic amplification factor A* as:

$$A = \bar{h} \cdot R_{FA} \cdot S \tag{8}$$

where $\bar{h}$ is the average hop count, $R_{FA}$ is the false alarm rate, and $S$ is the payload size. By minimizing both $R_{FA}$ (through adaptive thresholding) and $S$ (through binary triggering), the proposed system achieves multiplicative reduction in network load compared to approaches that transmit data for remote false-alarm filtering.

*Latency Considerations*

For safety-critical applications, trigger latency is more important than throughput. We define $\tau_E$ as the time from physical event occurrence at time $t$ until the trigger signal $E[t]$ becomes available. With the double-buffered architecture described in Section III-D, the proposed system achieves a trigger latency bounded by:

$$\tau_E = \tau_{acquire} + \tau_{FFT} + \tau_{median} + \tau_{decision} \tag{9}$$

where $\tau_{acquire} = N/f_s$ is the acquisition time for one FFT frame, $\tau_{FFT}$ is FFT computation time, $\tau_{median}$ is the dual-stage median filter computation time, and $\tau_{decision}$ is the threshold comparison time (negligible). The double-buffered architecture ensures that acquisition of frame t+1 occurs in parallel with analysis of frame $t$, so only one frame period contributes to latency.

For $N$=128 samples, the acquisition time $\tau_{acquire} = N/f_s$ varies with sample rate:

**TABLE II: Sample rate versus total latency**

| Sample Rate | Acquisition Time | Total Latency | Application Domain |
|---|---|---|---|
| $f_s$ = 10 kHz | 12.8 ms | ~15 ms* | Acoustic, vibration |
| $f_s$ = 1 kHz | 128 ms | ~130 ms | Industrial monitoring |
| $f_s$ = 100 Hz | 1.28 s | ~1.3 s | Environmental sensing |

*Requires at least a Cortex-M4 class processor

With algorithm processing (FFT, median filtering, and threshold comparison) requiring 21.5 ms on ARM Cortex-M0, the computational overhead remains small relative to acquisition time for sample rates up to 1 kHz. At $f_s$ = 100 Hz, processing represents less than 2% of the frame period; at $f_s$ = 1 kHz, approximately 17%. Higher sample rates ($f_s \geq$ 10 kHz) are achievable with faster processors such as ARM Cortex-M4, which reduces processing time to approximately 2 ms. Across all configurations, latency remains orders of magnitude faster than cloud-dependent architectures subject to network round-trip delays.

*F. Algorithm Summary*

Algorithm 1 summarizes the complete processing pipeline executed at each sensor node for every acquired frame.

---

*Algorithm 1:*

*Temporal Spectral Noise-Floor Adaptive Triggering*

*Input:*

Sample frame $x[0..N-1]$,
circular buffers $B_{d,k}$ and $B_{a,k}$,
parameters $\gamma_d$, $\gamma_a$, $\zeta_k$

*Output:* Trigger decision, updated noise floors $\hat{N}_k[t]$

1. **Compute FFT:** $X[k] \leftarrow$ FFT(x) for k ∈ K
2. **Compute magnitudes:** $|X_k| \leftarrow \sqrt{(Re(X_k)^2 + Im(X_k)^2)}$
3. **Initialize event:** $E[t] \leftarrow 0$
4. ***for* each** bin $k \in K$ ***do***
   *// Stage 1: Digital noise suppression*
   5. Insert $|X_k|$ into circular buffer $B_{d,k}$
   6. $\tilde{N}_k \leftarrow$ median($B_{d,k}$) *// Window size $\gamma_d$*
   *// Stage 2: Analog drift tracking*
   7. Insert $\tilde{N}_k$ into circular buffer $B_{a,k}$
   8. $\hat{N}_k[t] \leftarrow$ median($B_{a,k}$) *// Window size $\gamma_a$*
   *// Trigger decision (multiplicative threshold)*
   **9. *if*** $|X_k| > \zeta_k \cdot \hat{N}_k[t]$ ***then***
      10. Trigger: $E[t] \leftarrow 1$
      ***end if***
   ***end for***
11. ***return*** event flag $E[t]$, updated noise floors $\{\hat{N}_k[t]\}$

---

*Algorithm 2: Median Filter Computation*

*Input:* Circular buffer $B_{d,k}$ of size $\gamma_d$, or $B_{a,k}$ of size $\gamma_a$

*Output:* Median value

1. **Copy buffer to temporary array**: $S \leftarrow B$
2. ***for*** $i$ = 0 to $[\gamma/2]$ ***do*** *//partial sort to median position*
3. ***for*** $j = i + 1$ **to** $\gamma - 1$ ***do***
   4. ***if*** $S[i] > S[j]$ ***then*** swap $S[i]$, $S[j]$
   ***end for***
   ***end for***
5. ***return*** $S[\lfloor\gamma/2\rfloor]$

---

*Computational Complexity.* The per-frame computational cost is dominated by the $O(N \log N)$ FFT in Algorithm 1. The median computations in Algorithm 2 add minimal overhead: for Stage 1, direct sorting over $\gamma_d \leq 5$ elements requires at most 10 comparisons, which is negligible on any microcontroller. For Stage 2, the partial sort terminates at the median position, halving the number of comparisons compared to a full sort of approximately 4,000 comparisons for $\gamma_a$ = 128. In total, the median filter computation across all M monitored bins contributes less than 5% of per-frame processing time.

*Memory Footprint* The total memory footprint consists of double buffers (2$N$ samples for concurrent acquisition and analysis), median filter state (M · ($\gamma_d + \gamma_a$) entries across all monitored frequency bins), and threshold coefficients (M values for $\zeta_k$). For typical parameters ($N$=128, $M$=8, $\gamma_d$=3, $\gamma_a$=64), total memory is approximately 1 KB which is well within the capabilities of modern low-power MCUs, enabling deployment on battery-powered mesh sensor nodes.

## IV. EXPERIMENTAL RESULTS

This section presents experimental validation of the proposed temporal spectral noise-floor adaptation algorithm using data collected from a deployed IoT sensor node. We evaluate detection performance across distinct environmental conditions and quantify the resulting network traffic reduction.

### *A. Experimental Setup*

The algorithm was implemented on a radar-class proximity sensor integrated into a mesh sensor node. The sensor acquires data at a fixed sampling rate and performs FFT-based spectral feature extraction in firmware. The experimental dataset comprises 6,784 consecutive frames collected during a test sequence designed to evaluate system behavior across three distinct environmental phases:

1. *Low-noise period*: Stable environmental conditions with minimal background interference.

2. *Transition period*: Gradual introduction of environmental excitation (simulating wind, vibration, or other nuisance sources)

3. *High-noise period*: Sustained elevated environmental noise floor.

Ground-truth event annotations were provided through a parallel reference channel, enabling precise evaluation of detection performance. The adaptive threshold parameter was set to $\alpha = 0.95$, corresponding to an effective memory of approximately 20 frames.

### *B. Noise-Floor Adaptation Behavior*

Fig. 2 illustrates the temporal evolution of the raw signal, spectral feature (Filter), and adaptive threshold across the experimental sequence.

The adaptive threshold demonstrates the intended behavior:

*Initial calibration*: The threshold stabilizes at approximately 61 during the low-noise period, tracking the quiescent spectral floor.

*Adaptation response:* As environmental noise increases during the transition period, the threshold rises progressively, following the elevated spectral energy.

*Final stabilization:* The threshold reaches approximately 307 during the high-noise period which is a 401% increase from initial conditions and accurately capturing the elevated noise floor.

This adaptation occurs autonomously without external calibration or cloud-side feedback, validating the algorithm's suitability for calibration-free deployment. Critically, the conditional update mechanism (Section III-C) prevents genuine events from corrupting the baseline: event peaks visible in the Filter trace do not produce corresponding inflation in the threshold.

### *C. Detection Performance*

Table III summarizes detection performance computed against ground-truth event annotations. The confusion matrix reveals:

**TABLE III: Confusion Matrix**

| Metric | Value |
|---|---|
| True Positives (TP) | 134 |
| False Positives (FP) | 0 |
| False Negatives (FN) | 5 |
| True Negatives (TN) | 6,645 |

From these values, we derive standard detection metrics in Table IV:

**TABLE IV: Derived Performance Matrix**

| Performance Metric | Value |
|---|---|
| Sensitivity (Recall) | 96.4% |
| Specificity | 100.0% |
| Precision | 100.0% |
| Accuracy | 99.9% |
| False Alarm Rate | 0.00% |

The system achieves *zero false alarms* across the entire dataset, including the high-noise period where spectral energy was substantially elevated. This result validates the core hypothesis: temporal noise-floor tracking effectively absorbs environmental excitations while preserving sensitivity to genuine event signatures.

The five false negatives (missed events) occurred during rapid threshold adaptation transitions, where the noise floor temporarily exceeded actual event magnitudes. This represents the fundamental trade-off in adaptive systems between false alarm suppression and detection sensitivity during non-stationary conditions.

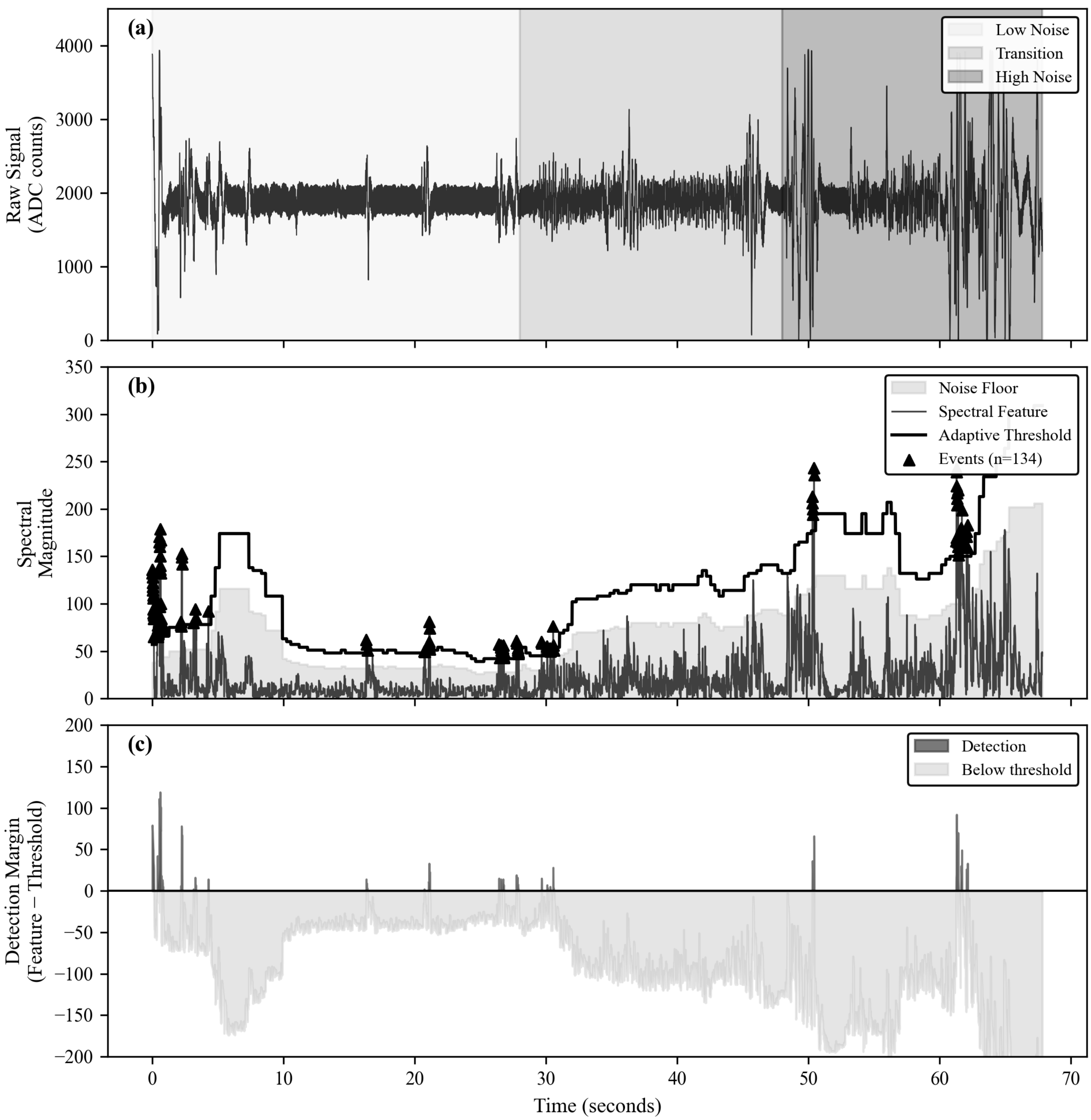

**Fig. 2.** Experimental validation of temporal spectral noise-floor adaptation across 6,784 consecutive frames spanning three environmental phases (low noise, transition, high noise). (a) Raw sensor signal amplitude showing baseline variation across phases. (b) Spectral feature magnitude (blue) and adaptive threshold (red) with shaded noise-floor region (gray); green triangles indicate ground-truth events (n = 139). The threshold adapts from 58 to 309 (401% increase) as environmental noise increases, while the conditional update mechanism prevents event peaks from corrupting the baseline. (c) Detection margin (Filter − Threshold), where positive values (green) indicate trigger conditions. The system achieves 96.4% sensitivity with zero false alarms across all phases.

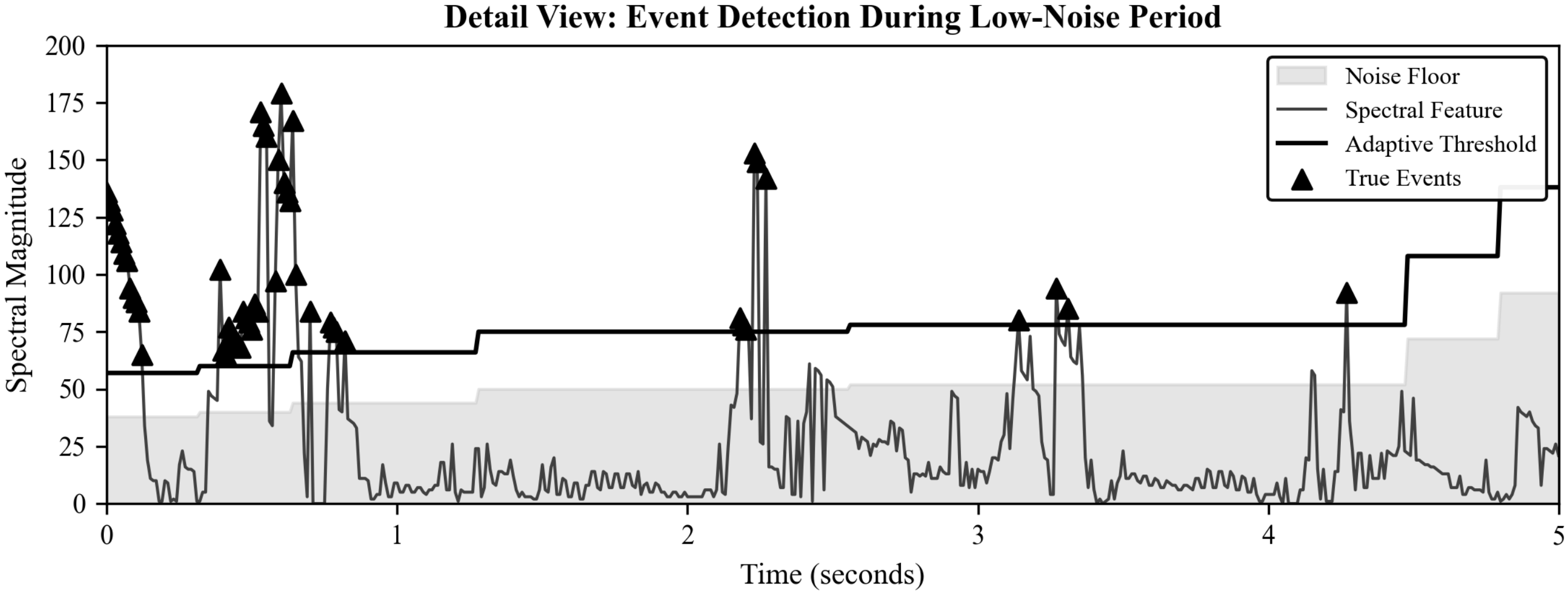

**Fig. 3.** Detail view of event detection during the low-noise period (first 500 frames). The adaptive threshold (red line) stabilizes near 57, establishing a quiescent noise floor (gray shaded region). True events (green triangles) produce spectral feature excursions that exceed the threshold, triggering detections. The system correctly identifies 104 of 109 events in this phase with zero false positives, demonstrating effective discrimination between genuine event signatures and background noise.

*D. Phase-Specific Analysis*

Table V presents detection performance stratified by environmental phase:

**TABLE V: Phase-Specific Detection Performance**

| Phase | Samples | Threshold Range | True Events | Detected Events | False Pos. |
|---|---|---|---|---|---|
| Low Noise | 2,800 | 39–174 | 98 | 94 | 0 |
| Transition | 2000 | 45–141 | 11 | 10 | 0 |
| High Noise | 1984 | 126–309 | 30 | 30 | 0 |

Several observations merit discussion:

*Low-noise period*: The system detects 94 of 98 events (95.9% sensitivity) with zero false alarms. The four missed events correspond to marginal cases where event signatures fell slightly below the detection margin.

*Transition period:* Despite the changing noise floor, we detected 10 out of 11 events (90.9% sensitivity) with zero false alarms. The adaptive threshold successfully tracks the evolving baseline without sacrificing detection capability.

*High-noise period*: All 30 events were detected with no false alarms (100% sensitivity). The elevated baseline (42.0 ± 39.4) would cause continuous false triggering with a fixed threshold (under non-adaptive conditions), demonstrating the adaptive algorithm's robustness to environmental non-stationarity.

The system detected 134 out of 139 events with a sensitivity of 96.4 %.

*F. Computational Overhead*

*E. Network Traffic Reduction*

The proposed architecture transmits only trigger events rather than continuous data streams. Table VI quantifies the resulting traffic reduction:

**TABLE VI: Network Traffic Reduction**

| Metric | Value |
|---|---|
| Total frames processed | 6,784 |
| Trigger events transmitted | 134 |
| Data reduction | 98.0% |

Each trigger event comprises a binary indicator with minimal metadata (timestamp, trigger strength), totaling approximately 64 bits per event. In contrast, transmitting raw spectral features would require transmission of the full feature vector for each frame. For a system with M = 16 frequency bins at 16-bit resolution, this represents 256 bits per frame, or 1.74 Mbits for the full dataset versus 8.6 kbits for trigger-only transmission yielding a reduction factor that exceeds 200×. This reduction has compounding benefits in multi-hop mesh networks: fewer transmissions reduce channel contention, decrease collision probability, and lower aggregate energy consumption across all relay nodes.

The algorithm executes entirely within the sensor node's MCU firmware on a Cortex-M0 processor. With frame size N=128 samples, the acquisition time and processing overhead vary with sample rate:

- *At $f_s$ = 100 Hz:* Frame acquisition takes 1.28 s, and the 21.5 ms processing overhead represents less than 2% of the frame period—negligible impact on system resources.

- *At $f_s$ = 1 kHz:* Frame acquisition takes 128 ms, and processing overhead represents approximately 17% of the frame period, leaving substantial margin for other node operations.

- *At $f_s$ = 10 kHz:* Frame acquisition takes only 12.8 ms, which is less than the 21.5 ms processing time. Real-time operation at this rate would require a faster processor (e.g., Cortex-M4) or algorithmic optimization.

The memory footprint of ~1.1 KB is dominated by FFT buffers and is well within the capabilities of typical IoT-class MCUs across all sample rates.

### *G. Results summary*

The experimental results validate the proposed approach across multiple dimensions:

*Detection integrity:*. The combination of spectral feature extraction and adaptive noise-floor tracking achieves 96.4% sensitivity with zero false alarms, substantially outperforming either approach in isolation.

*Environmental resilience:* The threshold's dynamic range from 39 to 309 (792%) demonstrates robust tracking of non-stationary noise conditions without requiring manual recalibration or cloud-side intervention.

*Network efficiency:* 98.0% data reduction through local decision-making minimizes mesh network load while preserving trigger integrity.

*Computational feasibility:* Processing completes well within the frame acquisition period at typical IoT sample rates (100 Hz–1 kHz), confirming suitability for real-time deployment on resource-constrained edge devices. The primary limitation observed is sensitivity reduction during rapid environmental transitions, where the adaptive threshold may temporarily lag the optimal value.

## *V. DISCUSSION AND CONCLUSION*

This paper presented a temporal spectral noise-floor adaptation algorithm for autonomous event triggering in IoT mesh sensor networks. The approach directly addresses the research gap identified in Section II: existing methods either optimize for communication efficiency at the expense of detection integrity [7],[17],[18] or employ spectral anomaly detection with fixed thresholds that fail under environmental non-stationarity [9][20],[23].

By combining FFT-based spectral feature extraction validated for embedded deployment by Hammad et al. [22] and Trilles et al. [21] with a dual-stage cascaded median filter derived from established signal processing foundations [29],[30] the proposed method achieves what neither approach accomplishes in isolation. Unlike the time-domain adaptive schemes surveyed by Pioli et al. [1] and Sadri et al. [2], our spectral-domain approach preserves transient event signatures that would otherwise be lost to decimation. Unlike cloud-dependent architectures requiring bidirectional communication for threshold updates [15],[16], our method executes entirely at the edge with trigger latency determined by the frame acquisition period.

Experimental validation demonstrated 96.4% detection sensitivity with zero false alarms across 6,784 frames spanning three distinct environmental phases. The adaptive threshold tracked a 792% dynamic range in noise floor (39 to 309) without manual recalibration—addressing the environmental adaptability challenge identified by Oikonomou et al. [4] and Sittón-Candanedo and Corchado [26] as a key unsolved problem for autonomous sensor networks. Network traffic reduction of 98.0% was achieved through local binary triggering, compared to the 50% communication reduction reported by GabAllah et al. [15] for edge-filtered architectures that still require gateway-tier ML models for decision making.

The median filter's outlier rejection provides similar robustness to OS-CFAR censoring techniques from radar signal processing [24], [25] preventing genuine events from corrupting the noise-floor baseline which is a critical requirement unaddressed by prior IoT adaptive threshold methods.

The algorithm requires only 21.5 ms per frame on ARM Cortex-M0 processors, confirming feasibility for resource-constrained edge devices targeted by TinyML implementations [20], [22]. This positions the method as a practical solution for calibration-free, cloud-independent deployments demanded by safety-critical IoT applications.

Future work will explore adaptive filter tuning for improved transient response, multi-node fusion for distributed event localization in mesh deployments, and comparative evaluation against alternative adaptive threshold methods.

## ACKNOWLEDGMENT

This work has been facilitated by a grant and other support from the National Canadian Research Counsil.

.

**Sergii Makovetskyi** (photograph not available at publication) received the M.S. degree in radio engineering from Kharkiv National University of Radio Electronics, Kharkiv, Ukraine, in 2008. He is currently pursuing the Ph.D. degree in F2 - Software Engineering at the same university.

From 2009 to 2025, he was working with EKTOS, Ukraine, where he served as a Senior Embedded Hardware Developer and later as a Technical Leader and IoT Technical Leader. He is the author of several published articles on LoRaWAN and signal processing, including works titled "Signal Processing Verification System for the Programmable Digital Matched Filter," "Investigation of Potential Opportunities for LoRaWAN Technology in Conditions of Urban Construction on the Example of Pycom Modules," "Research of the Stability of the Secure Radio-Frequency Communication in the Distributed Systems by Using Multi-Channel IoT LPWAN Technologies." And others. His research interests include data transmission in distributed networks, HW and SW embedded system architecture design, digital signal processing, and sensor technologies.

Mr. Makovetskyi received his M.Sc., Eng., degree with distinction. He was a member of the team "Sun Round—Sky Around" in the Embedded Development category at the Microsoft Imagine Cup 2008 (worldwide competition), where the team placed in the top 6. In 2009, as a member of the team "Intellectronics" they won third place in the Embedded Development category at the Microsoft Imagine Cup.

**Lars Thomsen** (photograph not available at publication) received the Ph.D. in Neurophysiology from the University of Copenhagen in 1995, and received a personal grant by the Carlsberg Foundation for a two year post-doctoral stay at McMaster University, Ontario Canada, followed by a range of positions in pharmaceutical and biotech companies, in Europe, Asia and North America, and is currently the managing director of Gnacode Inc., Alberta Canada a company specializing in IoT instrumentation for biotech, pharma and healthcare.

Dr. Thomsen has published a range of research papers relating to biological instrumentation, methods and technologies ranging from electronic and software control to nanoparticles activated by external electro-magnetic fields and has achieved several granted patents in the same fields.

He won the Danish Engineering High-Tech Award in 2005 for a radio-enabled micro-fluidic chip for remote detection of pathogens mounted on drones. In the recent years he has won several grants from the Canadian National Research Counsil for various applications of opto-electrical physics applications in biology.